\documentclass[]{pasj01}
\Received{}
\Accepted{2024/06/25}
 
\usepackage[switch,mathlines]{lineno}
\usepackage{longfigure}

\usepackage{dcolumn}

\def\Msun{$M_{\odot}$}
\begin{document} 
\renewcommand{\linenumbers}{}

\title{First VLBI Imaging of SiO $v=0$, $J=1\rightarrow 0$ Masers in VY Canis Majoris}

\author{
Hiroko \textsc{Shinnaga}\altaffilmark{1,2}, 
Miyako  \textsc{Oyadomari}\altaffilmark{1,2,3}, 
Hiroshi  \textsc{Imai}\altaffilmark{1,2,3}, 
Tomoaki \textsc{Oyama}\altaffilmark{4}, 
Mark J. \textsc{Claussen}\altaffilmark{5}, 
Masumi \textsc{Shimojo}\altaffilmark{6,7}, 
Satoshi \textsc{Yamamoto}\altaffilmark{8}, 
Anita M.S.  \textsc{Richards}\altaffilmark{9}, 
Sandra \textsc{Etoka}\altaffilmark{9}, 
Malcolm \textsc{Gray}\altaffilmark{10},
and 
Takeru \textsc{Suzuki}\altaffilmark{11}
}

\altaffiltext{1}{
Amanogawa Galaxy Astronomy Research Center, Graduate School of Science and Engineering, Kagoshima University,  \\
1-21-35 Korimoto, Kagoshima 890-0065, Japan}
 \email{shinnaga@sci.kagoshima-u.ac.jp, miyako.oyadomari@gmail.com, k3830453@kadai.jp, t.oyama@nao.ac.jp, mclausse@nrao.edu, masumi.shimojo@nao.ac.jp, yamamoto@phys.s.u-tokyo.ac.jp, a.m.s.richards@manchester.ac.uk, sandra.etoka@googlemail.com, malcolm@narit.or.th, stakeru@ea.c.u-tokyo.ac.jp}
 
 \altaffiltext{2}{
Department of Physics and Astronomy, Graduate School of Science and Engineering, Kagoshima University, \\
1-21-35 Korimoto, Kagoshima 890-0065, Japan}

\altaffiltext{3}{Center for General Education, Institute for Comprehensive Education, Kagoshima University,  \\ 
1-21-30 Korimoto, Kagoshima 890-0065, Japan}

\altaffiltext{4}{Mizusawa VLBI Observatory, National Astronomical Observatory of Japan, 2-12 Hoshigaoka, Mizusawa, Ohsyu, Iwate 023-0861, Japan}

\altaffiltext{5}{National Radio Astronomy Observatory, 1003 Lopezville Rd, Socorro, NM 87801, USA}

\altaffiltext{6}{Chile Observatory, National Astronomical Observatory of Japan, 2-21-1 Osawa, Mitaka, Tokyo 181-8588, Japan}

\altaffiltext{7}{Department of Astronomical Science, School of Physical Science, SOKENDAI (The Graduate University of Advanced Studies), 2-21-1 Osawa Mitaka, Tokyo, 181-8588, Japan}

\altaffiltext{8}{Department of Physics, The University of Tokyo, 7-3-1 Hongo Bunkyo-ku, Tokyo 113-0033, Japan}

\altaffiltext{9}{Jodrell Bank Centre for Astrophysics, School of Physics and Astronomy, Univ. of Manchester, Alan Turing Building, University of Manchester, M13 9PL UK}

\altaffiltext{10}{National Astronomical Research Institute of Thailand, 260 Moo 4, T. Donkaew, A. Maerim, Chiangmai 50180, Thailand}

\altaffiltext{11}{School of Arts and Sciences, University of Tokyo, 3-8-1 Komaba, Meguro, Tokyo 153-8902, Japan}


\maketitle

\begin{abstract}
 We achieved the first VLBI detections of the ground vibrational state ($v=0$) $^{28}$SiO (hereafter, SiO) and $^{29}$SiO masers of the $J=1\rightarrow 0$ rotational transitions, towards the 25 \Msun ~red supergiant (RSG) star, VY Canis Majoris (VY CMa), taking advantage of the high sensitivity of
the VLBI Exploration of Radio Astrometry (VERA) telescopes that coordinate with the Nobeyama 45 m telescope.  
In addition, we successfully detected the SiO $J=1\rightarrow 0$ transition in the $v=3$ state towards VY CMa for the first time with VLBI.  
The SiO $J=1\rightarrow 0$ maser spot in $v=0$ state was detected in the cross-power spectra taken with the baselines involving the Nobeyama 45-m telescope. 
The combination of previously reported absolute astrometry and the relative astrometry technique allowed us to derive the location of the SiO $v=0$ maser spot, {(RA, DEC) = ( 7${\rm ^h}$ 22${\rm ^m}$ 58.$^{\rm s}$32, $-$25$^{\circ}$ 46$^{\prime}$ 3.$^{\prime\prime}$4) in J2000 at an absolute positional accuracy of $\sim$100 milliarcseconds (mas).
The SiO $v=0$ maser spot is offset by the amount of ($\Delta$RA,  $\Delta$DEC)=($-$150, $-$300) (mas) to the southwest of the stellar position,
suggesting that the $v = 0$ maser spot is associated with its outflow activity.}
This observational study demonstrates that the brightest SiO $v=0$ maser spot is compact (3 mas), producing an extremely high brightness of $\sim$ 10$^7$ K. This indicates that the SiO $v=0$ maser action
may originate from strong shocks in the stellar wind emanating from this extreme RSG that leads to its intense mass ejection.
\end{abstract}

\maketitle
\pagewiselinenumbers

\section{Introduction}

{~~ Silicon monoxide (SiO) is one of the abundant molecules in the circumstellar envelopes (CSEs) of oxygen-rich evolved stars, such as red giant stars (RG) and supergiant stars (RSG), which are close to the end of their lives, generating the energy and materials they produced from nuclear fusion. }
They exhibit bright SiO (hereafter meaning $^{28}$SiO) maser emission over multiple vibrationally excited states ($v$ $\geq$ 0) in a radius of several au, or 2 $-$ 5 times the optical stellar radius of a long-period pulsating star, such as a Mira variable, and above the so-called ``molecular photosphere” within 1--2 times the stellar radius 
(e.g. \cite{gra09},  
\cite{tsu06} for the case of RSGs). Recent millimeter observations have detected such SiO masers in an extensive range of rotational-level quantum numbers, $J$, from which one can see the maser transitions, $J = n+1 \rightarrow n$ has been identified, where $n\geq0$. 
This suggests that these masers should trace the processes by which gases are ejected from the star by radial stellar pulsation and cooled down, leading to dust nucleation and condensation. Such dust particles are efficiently accelerated through their stellar radiative pressure; however, it remains unclear how asymmetry in the mass loss flow develops in the inner SiO maser spots cluster region to the outer H$_2$O maser spots cluster region in the CSE. In fact, SiO $v \geq$ 1 maser spots have been imaged intensively using the Very Long Baseline Interferometry (VLBI) technique.  \citet{ric16} reported detailed analysis of the results based on a sensitive VLBI imaging, using the VLBA, towards VY CMa, finding the extent of over 0.$^{\prime \prime}$1 from the star, corresponding to $\sim$ 110 au at a distance of 1.14 $^{+0.11 }_{-0.09}$ kpc \citep{cho08} and 1.20 $^{+0.13 }_{-0.10}$ kpc \citep{zha12}. 

On the other hand, the rotational transitions of SiO in its ground vibrational state ($v = 0$) generally exhibit weak emission (e.g. \citet{dev16}). Due to the low excitation temperatures of the rotational levels (e.g., 2 K and 6 K for the levels $J=$ 1 and 2, respectively), they are mostly thermally excited. 
However, some SiO emission sources have very bright and highly polarized features (e.g. \citet{cha95, tsu96, shi99, dev16}). Moreover, one considers SiO emission as probes to trace shocked regions caused by jets emanated by a young stellar object, interacting with the ambient gas cloud and SiO molecules through evaporation from dust particles via shock (e.g., \cite{ume92}). This phenomenon is observable in the shocks created by stellar winds interacting with the CSEs. 

VY Canis Majoris (VY CMa) is a RSG star about 25 times the mass of the Sun and one of the brightest stars in the infrared in the Milky Way Galaxy.  Due to its proximity (1.14 kpc or 1.2 kpc; \citet{cho08, zha12}) and high mass loss rate ( $6 \times 10^{-4}$ \Msun per year; \citet{she16}), the nature of the star has been studied in detail (e.g. \citet{smi01,sin23}). Owing to its high mass, this RSG is expected to collapse into a Type II core-collapse supernova or a stellar black hole (e.g. \citet{heg23}).  The CSE of VY CMa exhibits exceptionally intense SiO lines over many vibrational states \citep{cer93} along with strong water maser lines (e.g. \cite{ric14}).  

Detection of highly polarized emissions of the $J = 1 \rightarrow 0$ line \citep{shi99,shi17} indicates the presence of maser components. 
{ An interferometric mini survey of the SiO $v = 0$ rotational transition line emission was previously conducted with the Very Large Array (VLA) at the angular resolution of $\sim$ 0.\arcsec4--1.\arcsec1 towards six stars \citep{bob04} that did not include VY CMa. 
They found several velocity components of $v = 0$ maser towards all stars spread out, from a few times up to $\sim$ 10 times farther away compared with the extent of $v = 1$ maser components clustered (within 50 milliarcseconds, mas) near the central stars. Even with the higher angular resolution of the VLA observations, 0.\arcsec29 $\times$ 0.\arcsec12 \citep{shi17}, this was insufficient to distinguish compact maser components from extended thermal components unambiguously with the limited velocity resolution of 5.4 km s$^{-1}$.
\citet{shi17} found that three SiO clumps in the CSE of VY CMa show a strong Zeeman effect. Among them, Clump 2 is the closest to the central star, 138 au from the central star in the southwest direction. Their precise locations should provide important clues to the origin of the strong magnetic field indicated by a detection of the Zeeman effect \citep{shi17}.  

The VLBI technique gives us an extremely high angular resolution ($\ll$ 10 mas).  Detecting emission using the VLBI will prove the existence of very compact maser clumps.  Such compact maser clumps provide an essential clue into the nature of maser action 
in the SiO $v = 0$ $J = 1 \rightarrow 0$ line. 
This paper reports the first confirmed VLBI detection of SiO and $^{29}$SiO $v=0$ , $J = 1 \rightarrow 0$ emission toward VY CMa at an extremely high angular resolution of $\sim$ 1 mas.   

\section{Observations and data reduction}
\label{sec:observations}
~~~ We conducted VLBI observations of SiO masers in the rotational transition of $J=1\rightarrow 0$ on the vibrational levels of $v=$ 0, 1, 2, and 3 (at 43.423798, 43.122027, 42.820539, and 42.519340 GHz, respectively) and $^{29}$SiO $J=1\rightarrow 0$ masers at $v=0$ (42.879850 GHz; Table 1) towards VY CMa using the four telescopes of the VLBI Exploration of Radio Astrometry (VERA) array 
and the 45 m telescope of the Nobeyama Radio Observatory (NRO) on February 8 and March 8, 2018.  {  This paper reports the results and discussion based on the first successful 
VLBI session. }

VY CMa, its calibrator, J071814.2$-$181304, and fringe finder, J073019.1$-$114113 were observed for a total of 6.7 hours. In all the stations, the left-hand circular polarization signals were filtered into two base band channels (BBCs), each with a bandwidth of 512 MHz, to cover a total bandwidth of 1024 MHz, so that it includes all of the SiO line frequencies mentioned above.  The frequency setting of the observation was similar to that of the observations reported by \citet{oya16} and \citet{oya18}. The filtered signals were digitally sampled at a rate of 4~Gbps in 2-bit quantization using ADS3000+ and finally recorded using the OCTADISK data storage system \citep{oya16}. 

{In VERA, the dual beam system was used in parallel for high-precision astrometry, which is necessary for the accurate registration of different maser line maps on a common coordinate system. 
The dual beam system 
was being setup for the VERA astrometry as described in \citet{ima12a}. }


One of the dual beams was pointed toward VY CMa, 
for which only the signals of the SiO $v=2$ and $v=3$ maser lines were transferred for astrometry, while the other beam was pointed toward the position reference source J072524.4$-$264033 (hereafter abbreviated as J0725), 1.06$^{\circ}$ away from VY CMa. The transferred signals were sampled using a pair of ADS1000 in 2-bit quantization, filtered onto 16 BBCs each with a bandwidth of 16 MHz, two BBCs for the two ($v=2$ and 3) maser lines, and other BBCs for the reference source continuum emission, and recorded onto another OCTADISK system at a rate of 1~Gbps. 


The recorded signal data were correlated using the software correlator in the Mizusawa VLBI Observatory. For the 4~Gbps data, two types of correlated data were produced. The first one had five spectral windows from two BBCs for the five SiO maser lines mentioned above, each with a frequency coverage of 16~MHz and 1024 spectral channels, yielding a velocity channel spacing of $\sim$ 0.11 km~s$^{-1}$. The other had two spectral windows for the continuum calibrator and the fringe finder, each with a frequency coverage of 512 MHz and 512 spectral channels. 


Two types of correlation data were produced for the 1~Gbps data as well; the one including two BBCs for the SiO $v=2$ and 3 maser lines each with 1024 spectral channels and {the other including the 16-MHz BBCs for 
the continuum sources each with 16 spectral channels. } An accumulation period of 1 sec was commonly adopted in all datasets. 

Data calibration and image cube synthesis were performed using the Astronomical Image Processing System (AIPS) of the National Radio Astronomy Observatory (NRAO), with Python/ParselTongue scripts\footnote
{See ParselTonuge wiki: http:\/\/www.jive.nl\/jivewiki\/doku.php?id=parseltongue:parseltongue.} for automated processing \citep{ima20}. 

The fringe visibility amplitudes were calibrated using tables of antenna gains and system noise temperatures. Because these parameters were commonly given to all spectral windows spread over a width of $\sim$1~GHz and the bandpass characteristics are affected by the wide band electronics and bright maser lines, the absolute flux density scales is 
ambiguous (uncertainty of up to 50\%) among the spectral windows. 

Instrumental delays, delay rates, and phase offsets of the fringe visibilities were calibrated using the broad-band data of the continuum calibrators, including the fringe finder. Fringe fitting and self-calibration were made using the spectral channel data including bright maser emission in the individual SiO maser lines and the calibration solutions were applied to other channels in the same maser lines. These standard procedures were made to obtain image cubes for the individual maser lines with a quality better than that in the phase-referencing method mentioned later. 

Image cubes of the SiO maser emissions were created using the CLEAN deconvolution algorithm with naturally weighted visibilities. Table \ref{tab:maps} provides the parameters of the Gaussian synthesized beams and the 1-$\sigma$ noise levels (in the maser emission-free channels) of the individual maser line maps. 

The data from VERA's dual beams alone could detect only SiO $v=2$ $J=1\rightarrow 0$ maser emission, resulting in the high-precision astrometry being unsuccessful. 
{The flow of the data reduction steps for the astrometry is described in detail by \citet{ima12a,ima12b}.} The solutions of fringe fitting and self-calibration were obtained with the data including the bright $v=2$ maser emission and applied to the data of the position-reference source J0725, whose absolute coordinates have been accurately determined ($\sigma<$1~mas)\footnote{see: http:\/\/astrogeo.org\/rfc\/.}. Unfortunately, as mentioned above, we could not detect J0725 even in the phase-referenced image or
determine the absolute coordinates of the bright maser spot with respect to J0725 because it 
had a flux density of only 30~mJy at 43~GHz \citep{zha12}, which is too faint to be detected at a low elevation with limited recorded signal bandwidth.

{
Instead, we adopted the coordinates of the SiO $v=$1 maser spot determined by \citet{zha12} as the absolute coordinates of the $v=1$ and $v=2$ maser spots at the local standard of rest (LSR) velocity of 17.9 km s$^{-1}$ detected in our observation based on their VLBA astrometry: (RA, DEC) = (7$^{\rm h}$ 22$^{\rm m}$ 58.$^{\rm s}$3283, -25$^{\circ}$ 46$^{\prime}$ 3.$^{\prime\prime}$075) in J2000. 
Some discrepancy can be expected between the positions of the SiO 
maser spots detected in this work and Zhang et al. (2012). 
We speculate that such a discrepancy, corresponding to the former spot's absolute position uncertainty, is around 60 $-$ 100 mas, taking into account the star's proper motion and the extent of the maser spot distribution (within 60 mas). 
In fact, \citet{zha12} derived the stellar proper motion of (-2.8, 2.6) mas yr$^{-1}$, resulting in (-34, +31) mas in the difference over 12 years (from 2006 until 2018).  Note that \citet{zha12} reported the 10 mas uncertainty in their stellar position as origin.  \citet{}
}

For relative astrometry, the calibration solutions obtained with the SiO $v=1$ maser spot mentioned above were commonly applied to all the maser line data. 
With the exception 
of the SiO $v=0$ maser, for which the image cube was obtained only with the self-calibration procedures, 
these calibration solutions were transferred to the data of $v=1$ maser 
to facilitate relative astrometry. 
{
Following the map registration described in detail in \citet{ima10} and \citet{ima12a}, 
all maser line maps were registered in a common coordinate system with a typical relative accuracy of $\sim$0.7~mas.}
Fringe phase rotation for setting the position of the $v=0$ maser at the phase-tracking center of the visibility data would not be helpful for detecting the $v=0$ maser in the phase referencing because of too large uncertainty in the maser position as to be mentioned in the next section. 



\begin{table}[th]
\caption{Parameters of the synthesized maser line maps.}\label{tab:maps}
\begin{tabular}{rr@{$\times$}lcr}\hline\hline
Maser line  (GHz) & \multicolumn{3}{c}{Gaussian beam parameters} & \multicolumn{1}{l}{1-$\sigma$} \\
($J=1\rightarrow 0$) & $\theta_{\rm maj}$ & $\theta_{\rm min} {\rm [mas]}$ & P.A.[$^\circ$] & noise [Jy] \\ \hline
SiO $v=0$ ~43.423798 & 1.39 & 1.06 & $-$27.9 & 0.23\\
SiO $v=1$ ~43.122027& 1.21 & 0.84 & $-$11.9 & 0.59\\
SiO $v=2$ ~42.820539 & 1.19 & 0.93 & 6.8 & 0.25\\
SiO $v=3$ ~42.519340& 1.38 & 0.49$^{\ddag}$ & 16.3 & 0.57\\
$^{29}$SiO $v=0$ ~42.879850 & 1.47 & 1.33 & $-$22.54 & 0.18\\ 
\hline
\end{tabular}
\noindent
$^{*}$Frequencies were obtained from the Lovas catalogue (https://physics.nist.gov/cgi-bin/micro/table5/start.pl). \\
$^{\ddag}$Beams narrower than those in other maser lines are attributed to the visibility data from the baselines including the Nobeyama 45 m telescope, 
for which the recorded signals were seriously attenuated. 
\end{table}
\clearpage
\begin{figure}[h]
\includegraphics[width=8.5cm]{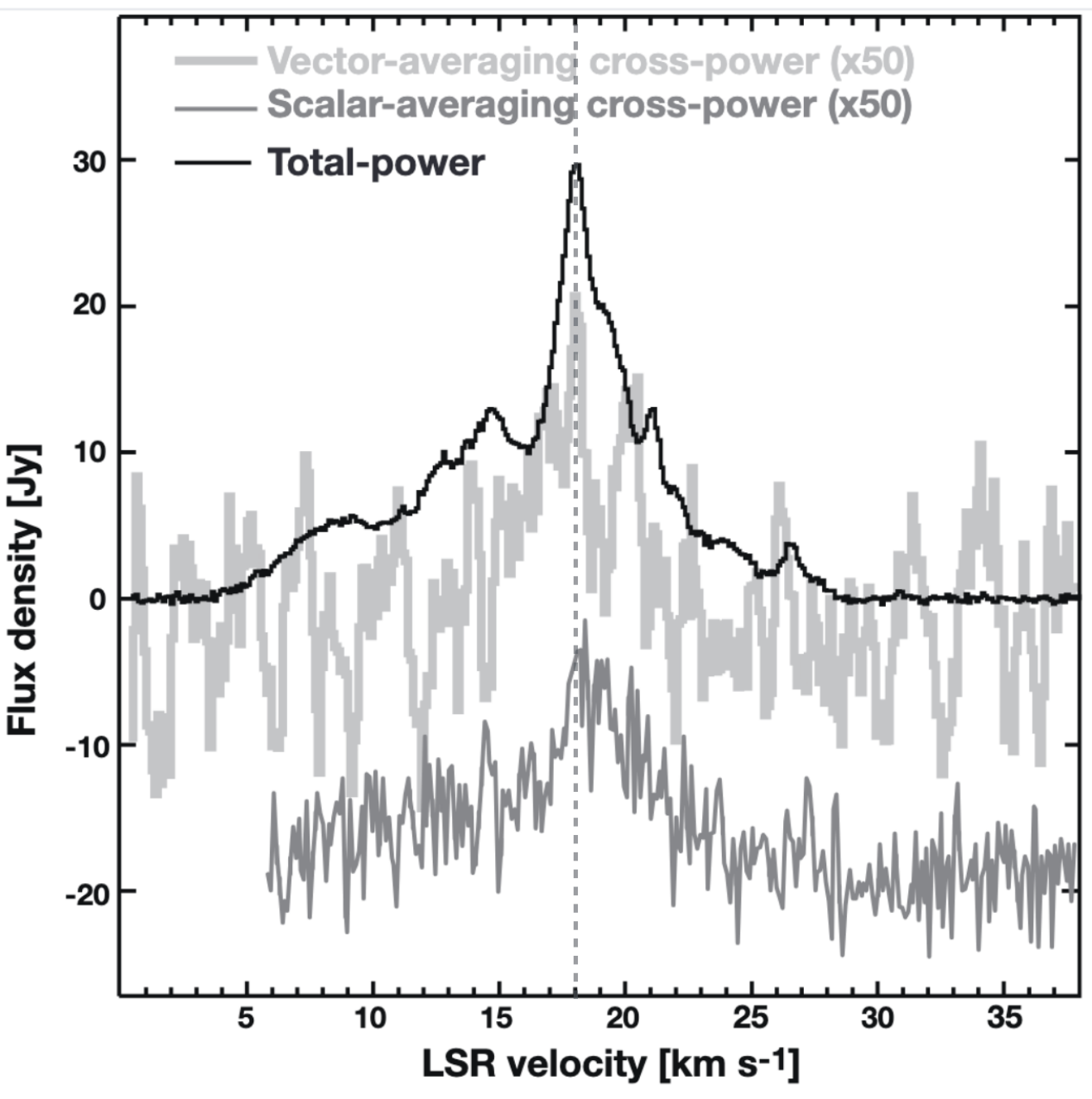} 
\caption{Spectra of the SiO $v=0$, $J=1\rightarrow 0$ line toward VY CMa. The total-power spectrum taken with the NRO 45 m telescope is plotted as a black thin line. The spectral baseline was subtracted. Out of the scalar-averaging cross-power spectra exhibiting the maser detection, only that taken using the VERA Iriki–-NRO 45 m baseline is plotted as a thin gray line 
with the spectral baseline offset for clarity. The vector-averaging cross-power spectrum was obtained using a combination of the baselines including the NRO 45 m telescope and is plotted in a thick gray line. Cross-power spectra are magnified by a factor of 50 in the flux density scale.The vertical dotted line corresponds to $V_{LSR}$= 17.9 km s$^{-1}$ where the $^{28}$SiO $v=0$ maser emission peaks.}
\label{fig:v=0_spectrum}
\end{figure}
\clearpage
\begin{figure}[h]
\includegraphics[width=16cm]{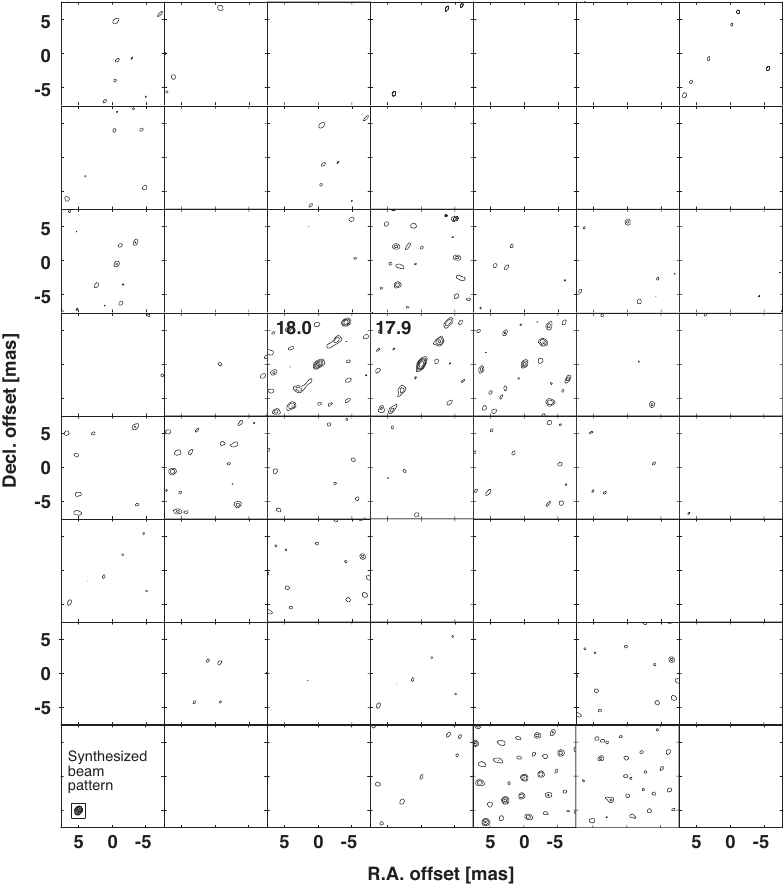} 
\caption{Channel maps are presented for the SiO $v=0$ $J=1\rightarrow 0$ maser emission toward VY CMa obtained after fringe fitting and self-calibration procedures and displayed around the velocity channels for which the maser emission is detected. Each channel map shows every velocity spacing of 0.11~km~s$^{-1}$ 
at contour levels of 
0.62, 0.83, 1.04, 1.25, and 1.66 Jy~beam$^{-1}$, corresponding to signal-to-noise ratios of 3, 4, 5, 6, and 8, respectively. 
The LSR velocities are labeled in the two sub-panels at 17.9~km~s$^{-1}$, in which the maser is detected above an 8~$\sigma$ noise level, and 18.0~km~s$^{-1}$,  
in which the maser detection is less significant (at a 6~$\sigma$ noise level) but found at the same position as that of the confident maser detection at 17.9~km~s$^{-1}$.
The synthesized beam size is drawn in the panel at bottom left corner. 
} 
\label{fig:v=0_channelmap}
\end{figure}
\clearpage
\begin{figure}[h]
\includegraphics[width=8.5cm]{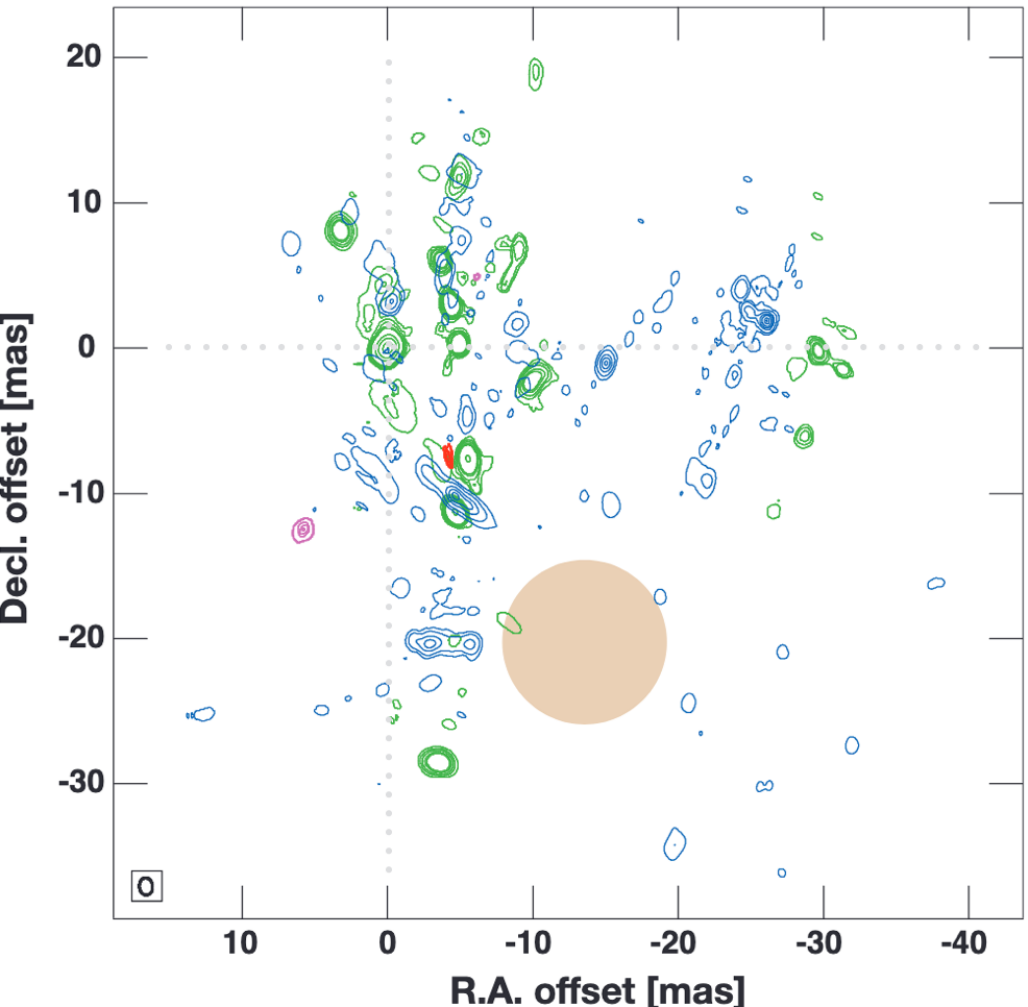}
\caption{{Composite map of SiO $J=1\rightarrow 0$ maser spots in $v=$1 (blue), 2 (green), 3 (red) states and $^{29}$SiO $v=0$ state (magenta) surrounding VY CMa. 
 The origin of the map is set to the position of the brightest SiO $v=2$ maser spot 
at $V_{\rm LSR}=$17.9~km~s$^{-1}$ used as fringe-phase and position reference in the final calibration process for finer imaging. 
The position of the referenced maser spot is estimated to be at 
(RA, DEC) = (7$^{\rm h}$ 22$^{1\rm m}$ 58.$^{\rm s}$328, -25$^{\circ}$ 46$^{\prime}$ 3.$^{\prime\prime}$08 ) in J2000.  
 Note that the phase referencing using the brightest $v=2$ maser spot was also performed to  obtain this composite map with a higher sensitivity than that obtained from the fainter $v=1$ maser spot.
}
The brown filled circle represents the angular size of the stellar photosphere (11.3~mas, \cite{wit12}) or a diameter of 14 au at the distance of the star. 
{ The stellar position in this diagram suffers from significant uncertainty, up to $\sim$10~mas (please find the details in the main text).} 
The synthesized beam of the observation is shown at the bottom left corner of the panel.  
}
\label{fig:map}
\ \\
\includegraphics[width=8cm]{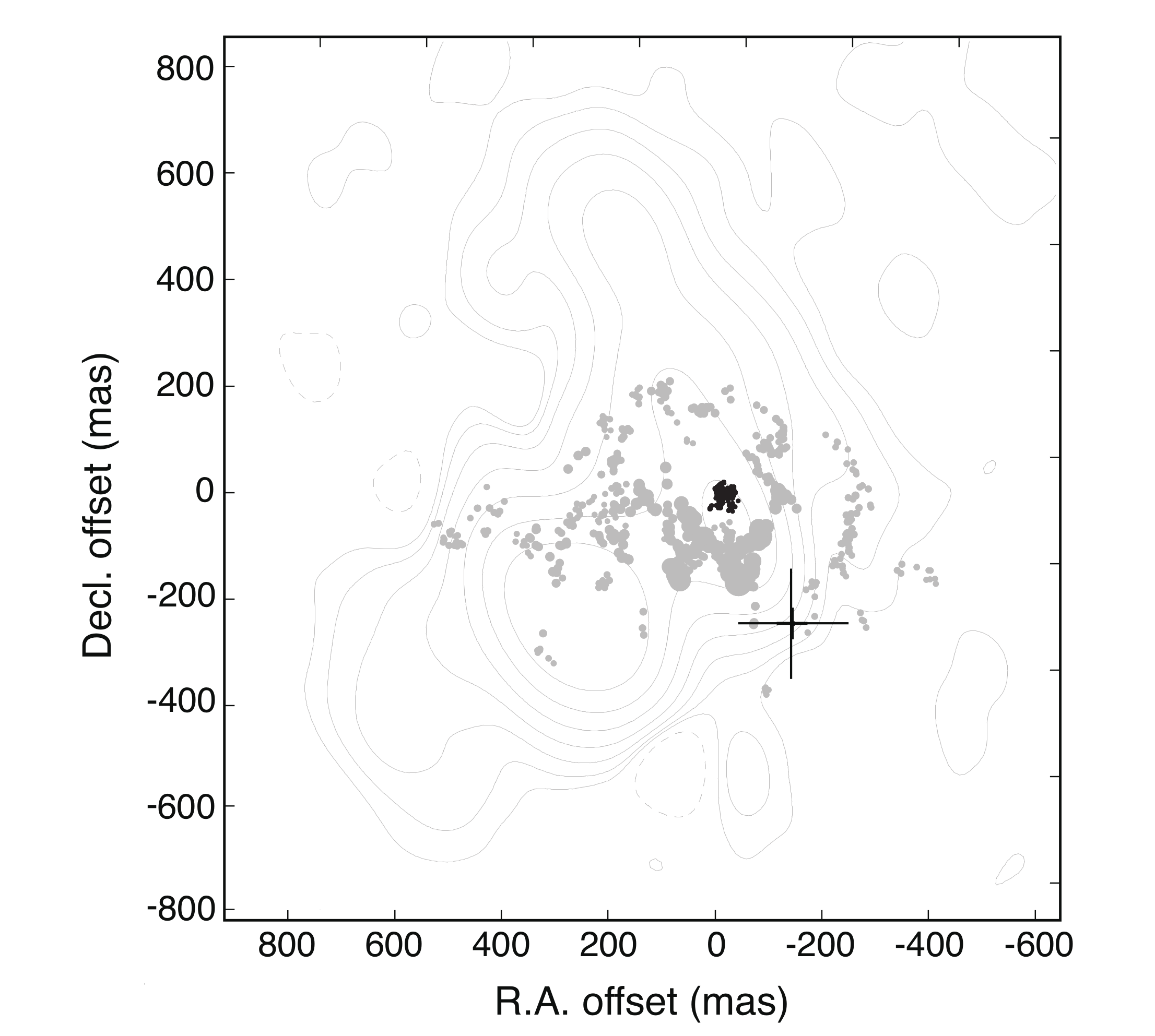} 
\caption{The big '+' sign represents the location of the SiO $v=0$, $J=1\rightarrow 0$ maser spot that we detected with the VERA + Nobeyama 45 m telescope, 
overlaid on the $\nu =$ 321 GHz continuum emission map, shown in the gray contours and the H$_2$O maser spots of the $J_{Ka, Kc}= 10_{2,9} - 9_{3,6}$ transition at $\nu=$ 321.22564 GHz in gray dots \citep{ric14} in the CSE of VY CMa.  The size of the symbol 
(100 mas) corresponds to the uncertainty in the position of the $v=0$ maser spot with respect to the brightest SiO $v=1$, $J=1\rightarrow 0$ maser spot. 
The black dots denote the locations of SiO $v \ge 1$ maser spots  (presented in Fig.\ \ref{fig:map}).  { The map origin is the same as Fig. 3. }
}
\label{fig:v=0_map}
\end{figure}

\section{Results}
\label{sec:results}
~~~ Fig.\ \ref{fig:v=0_spectrum} represents the total- and cross-power spectra of the SiO $v=0$  $J=1\rightarrow 0$ maser line toward VY CMa. The total-power spectrum shows that the emission line covers a wide velocity range 
(from $V_{\rm LSR}=$4 to 28~km~s$^{-1}$). The scalar-averaging cross-power spectrum shows the evidence of VLBI detection of the maser line with a cross-correlated flux density much lower than the total-power flux density in a velocity range of 15--25~km~s$^{-1}$ only in the baselines including the NRO 45~m telescope, which has 
high sensitivity and short baselines. 
It indicates that the maser emission is spatially significantly resolved, although it has a velocity width comparable to that seen in the total-power spectrum. {The vector-averaging cross-power spectrum obtained in the combination of the baselines of the VERA 
and 
the NRO 45~m telescopes also indicates a tentative detection of the maser line. We achieved the resultant angular resolution of 1 mas (Table 1). }Note that the correlated flux density appears to be reduced due to some coherence loss over long integration, even with self-calibration. 

Despite the tentative maser detection, fringe fitting and self-calibration using the detected maser emission at $V_{\rm LSR}=$17.9~km~s$^{-1}$ were conducted successfully, enabling us to synthesize the maser maps in two consecutive velocity channels including the phase-reference channel down to an 8-$\sigma$ noise level. Fig.\  \ref{fig:v=0_channelmap} shows the channel maps of the maser, including the channels with null detections ($<6\sigma$), for comparison. Unfortunately, we were unable to map 
most of the velocity components visible in the scalar-averaging cross-power spectrum. It is difficult to yield such a high-sensitivity map given the 
limited quality of the data calibration solutions. 

Fig.\ \ref{fig:map} shows the registered map of the maser spots of SiO $v=$1, 2, and 3,  $J=1\rightarrow 0$ and $^{29}$SiO $v=0$,  $J=1\rightarrow 0$ transitions. 
{The origin of the map of Fig.\ \ref{fig:map} is set to the position of the brightest SiO $v=2$ maser spot at $V_{\rm LSR}=$17.9~km~s$^{-1}$ used as fringe-phase and position reference. 
As mentioned in Sect. 2, the referenced maser spot is assumed to be located at around (RA, DEC) = ( 7${\rm ^h}$ 22${\rm ^m}$ 58.$^{\rm s}$33, $-$25$^{\circ}$ 46$^{\prime}$ 3.$^{\prime\prime}$1) in J2000 \citep{zha12}.  Note that phase referencing using the brightest $v=2$ maser spot was also performed in order to obtain this composite map with a higher sensitivity than that obtained from the fainter $v=1$ maser spot.} 
{The overall distributions of the SiO $v = 1$ and 2, $J=1 \rightarrow 0$ maser spots are consistent with those observed in 2007 \citep{ric16}.  One can recognize the features of chains of maser spots extending northeast and east from the central star, similarly to F1 in Fig.~2 of \citet{ric16}. 
It is possible to estimate the stellar position using chains of maser spots that look radially aligned from the central star (e.g., \cite{zha12,ric16}).  
\citet{zha12} adopted this method and confirmed that the estimated stellar positions in the maser maps are consistent within $\sim$10~mas. 
Thanks to our observations of the bright $v=1$ and $v=2$ SiO maser components, one can find a few structures that seem radially aligned in the present study, similar to those of \citet{zha12}. On the other hand, this method causes some ambiguities in the identification of the radial structures without knowing the true motions and physical connections between the 
maser features. 
 Considering the risk in drawing lines by connecting maser spots to find such radially-aligned structures and possible deviations from the linear distributions caused by radial expansion of stellar outflow, the estimated stellar position would have an uncertainty of up to $\sim$10~mas.}
{ The brown circle in Fig.\ \ref{fig:map} represents the stellar diameter, the central position of which was inferred applying the above-mentioned discussion. 
The angular size of the circle represents the stellar photosphere, measured to be 11.3±0.3 mas \citep{wit12}.}
{ The central position of the brown circle in Fig.\ \ref{fig:map} is shifted roughly ($\Delta$RA, $\Delta$DEC) = ($-$13 mas, $-$20 mas) from the brightest $v=$2 maser spot.  Due to the difficulties in astrometry described in Section 2, we cannot yield the absolute position of the star itself with this data set.   
} 

{ In addition, this 
study also represents the first VLBI detections of the SiO $v=3$ and the 
SiO and $^{29}$SiO $v=0$ maser emission 
towards VY CMa. } As suggested by \citet{ima10} and \citet{oya18}, the SiO $v=3$ masers are closely associated with the $v=1$ and $v=2$ masers. { On the contrary, the $^{29}$SiO $v=0$ maser spot seems to have a larger offset from the $v=1$ and $v=2$ maser spots.  
However, the projected distance to the $^{29}$SiO $v=0$ maser spot from the central star is comparable to  
those of the $v=1$ and $v=2$ maser spots. 
}

Fig.\ \ref{fig:v=0_map} shows the location of the detected SiO $v=0$ maser spot with respect to other brighter SiO $v\geq 1$ masers. The maps of SiO masers in the other transitions  
are superimposed on the diagram shown in the left panel of Figure 3 of \citet{ric14}, based on the estimated position of the star. { The relative location of the $v=0$ maser spot with respect to other masers has a large uncertainty, up to $\sim$ 100 
mas. 
Nevertheless, this uncertainty does not affect our scientific discussion, as described in the following.  }

{
We find that the position of the $v=0$ maser spot was shifted by an amount of ($\Delta$ RA, $\Delta$ DEC) = ($-$150, $-$300) mas with respect to the $v=1$ bright maser spot mentioned above.   
The estimated position of the $v=0$ maser spot is at  (RA, DEC) = ( 7${\rm ^h}$ 22${\rm ^m}$ 58.$^{\rm s}$32, $-$25$^{\circ}$ 46$^{\prime}$ 3.$^{\prime\prime}$4) in J2000 with an absolute positional accuracy of $\sim$100~mas. In principle, one can determine the relative positions of a faint maser spot with respect to a bright maser spot used for data calibration as a fringe-phase reference. 
In reality, we could not detect the faint maser spot in this phase-referencing technique. 
Instead, we applied the data calibration solutions inversely obtained from the data of the $v=0$ maser to the $v=1$ masers. In this scheme, we were able to detect several candidates of bright, but 
defocused, 
$v=1$ maser spots whose relative positions are consistent with those found in Fig.\ \ref{fig:map}, but with large uncertainties as noted 
above due to the tentative detection 
in the false brightness peaks. 
Thus, the uncertainty in the position of the $v=0$ maser spot is dominated by the error in the relative position with respect to the referenced $v=1$ maser spot and the absolute position error of this referenced maser spot. 
}

{ The estimated location of the VERA-NRO 45m detected SiO $v=0$ maser spot is further 
from the central star than other SiO $v \ge 1$ masers and bright 22 and 321 GHz H$_2$O masers in the $v=0$ state and 658 GHz water maser spots in the $v=1$ state observed with the VERA, VLBA and ALMA, respectively (within 150 mas in radius,  e.g., \cite{cho08,ric14,ric16}), while the 
SiO $v=0$ maser spot resides within the range of extended water maser spots at 321, 325, and 658 GHz (within 500--700~mas,  \cite{ric14}) as well as within the continuum components traced at 321, 325 and 658 GHz \citep{ric14}.  }
{ Among the SiO $v=0$ $J=1\rightarrow 0$ maser clumps mapped using the VLA \citep{shi17}, Clump 2 is the closest to the location of the VERA-NRO 45m detected SiO $v=0$ maser spot, among the three clumps,  but not on the exact position.  
The SiO $v=0$  $J=1\rightarrow 0$ emission 
consists of a mixture of maser and thermal emission. 
In addition, it is worth noting that the location of the 
SiO $v=0$ maser spot is in the direction of the "SW Clump," which is visible in the HST/F1042M image and the ALMA Band-7 image with an extended beam of 0$^{\prime\prime}$\hspace{-2pt}.9 \citep{kam19}. 
The VERA-NRO 45m SiO $v=0$ maser spot appears to be detected at almost the same position as one of the OH 1612 MHz transition maser spots which will be described in a separate paper (Shinnaga et al. in preparation).  
The VERA-NRO 45m SiO $v=0$ maser spot is likely excited by the active mass loss processes of the RSG and originate from some shocks in the CSE.  } 

\section{Discussion}
\label{sec:discussion}

~~~ The first VLBI detection of the SiO $v=0$ $J=1\rightarrow 0$ maser provides a significant impact on the astrophysical interpretation of this line emission. This line has been considered 
to be a shock tracer in outflows found not only in star forming regions \citep{ume92} but also in the CSEs (e.g., \cite{shi17}).  

The scalar-averaging cross-power flux density of the detected SiO $v=0$ maser spots has been measured to be $\sim$0.3 Jy, i.e., roughly equal to its vector-averaging cross-power flux density. This should impose 
a lower limit to the maser flux density because the self-calibrated image of the maser spot shows a higher flux density ($\sim$1.7~Jy), and gives the following lower limit to the maser brightness: 
{  
\begin{equation}
T_{\rm b} = \frac{\lambda^{2} S_{\nu}}{2 k\Omega_{\rm beam}}\approx \\ 1.5 \times10^7\frac{[(\lambda_{\rm SiO v=0}/6.95 {\rm mm})^2 ] }{[\Omega_{\rm beam}/{\rm 3~mas}]}\frac{{[S_{\nu}]}}{{{\rm 0.3 Jy}}}\;\; {\rm [K].}
\end{equation}
}
\noindent
{ The estimated extremely high brightness temperature of $1.5 \times 10^7$ K for 
SiO $v=0$ $J=1-0$ transition 
very strongly suggests the emission originates from maser action.  }This limit is two orders of magnitude higher than that given for a maser spot unresolved with the VLA ($T_{\rm b}\geq 10^5$~K, \cite{bob04}), but corresponds to 
the lowest class of SiO maser brightness (typically $T_{\rm b}\sim 10^9$~K for $v=1$ and $v=2$ masers). 

The location of the SiO $v=0$ maser with respect to the central star implies that this maser spot is associated with one side of the outflow converging in the northeast -- southwest direction. The overall brightness distributions of the SiO $v=1$ and $v=2$ masers are 
approximately consistent with those found in \citet{cho08}, \cite{zha12}, and \citet{ric16} and are biased in three directions: north, northwest, and southeast of the estimated stellar position. The origin of these persistent biased distributions over ten years is unclear, although the relevant spatial-kinematics has been investigated in detail on larger scales { (e.g., \cite{hum07,hum21,hum24})}. We can recognize the distribution of 6$_{1,6}$ $\rightarrow 5_{2,3}$ in $v=0$ state 22 GHz H$_2$O masers 
distributed in similar directions \citep{cho08}. 

{
The SW clump that was first identified in the HST images 
\citep{smi01} is of particular interest as in the southwest direction, metallic species such as TiO$_2$ \citep{bec15}, NaCl \citep{dec16} and PN \citep{rav24} are also detected, indicating some special condition is met to create these rare species.  The SW clump is within the extended H$_2$O 183 GHz 
$v=0$ 3$_{1,3}$ $\rightarrow 2_{2,0}$ maser clumps extending in south and southwest directions \citep{ric18}. Among them, quite a few H$_2$O 183 GHz $v=0$ maser clumps are located inside the SW clump.  } 
The SiO $v=0$ maser spot seems to be located on the downstream of the same side of the outflow as that of those masers. Although the SiO $v=0$ maser spot is still located within the dust continuum emission region, a steep gradient of the continuum emission brightness is visible \citep{ric14}. 

Revisiting the shock hypothesis \citep{ume92}, it is expected that SiO molecules are well released by the evaporation of dust, yielding the SiO $v=0$ maser excitation at a large distance ($\sim$280 AU) from the star. 

\citet{shi17} discussed the origin of apparent strong magnetic fields in the regions of SiO $v=0$ emission. Based on the high
degree of linear polarization, the estimated lower limit to the field strength, $\sim$10~G, is one order of magnitude larger than that found in SiO $v=1$ $J=1\rightarrow 0$ maser spots around the stellar surface (e.g., \cite{ric16}). However, they also claimed the possibility that the magnetic field strength implied for the SiO $v=1$ masers without measurement of circular polarization is possibly 
underestimated. On the other hand, the derived strong magnetic field is expected to strongly regulate the spatio-kinematics of the outflow, which appears to be inconsistent with the complexity of the outflow observed in a wide variety of radio emission. In fact, a simple energetics comparison between the magnetic field and the kinematics of the outflow exciting the maser is as follows: 

\begin{equation}
\frac{E_{\rm magnetic}}{E_{\rm kinematic}} = 600 
\frac{[B/{\rm 10~G}]^2}
{([n_{{\rm H}_2}/{\rm 10}^9{\rm ~cm}^{-3}][v/{\rm 20~km~s}^{-1}]^2)}, 
\end{equation}

\noindent
This estimate strongly suggests that the outflow should be controlled by a magneto-hydrodynamical force. In fact, ALMA polarimetric observations \citep{vle17} yield a much lower value for the lower limit of the magnetic field strength ($>$13~mG) at higher rotational transitions of SiO masers. An alternative interpretation for the magnetic field derived by \citet{shi17} is that the magnetic field may be enhanced in the maser region that may be increased to have a much higher density than that of the average outflow. 
Maser detection reported in this paper provide some support for this possibility.  

\section{Conclusions}

~~~~~{ We successfully detected SiO and $^{29}$SiO $v=0$ maser components with the VLBI for the first time.  
In addition, we achieved the detection of $J=1\rightarrow 0$ transition in the $v=3$ state towards VY CMa for the first time with VLBI.}  The spatial location of the detected SiO $v=0$ maser component strongly suggests the masing phenomena is closely linked to the intense mass loss processes originating from the RSG star. \par
To gain deeper insights into the underlying mechanism of maser phenomena associated with the circumstellar magnetic field, we stress the importance of the future observational studies measuring the circumstellar magnetic field, in conjunction with related theoretical studies.  These combined effort may provide important clues toward a comprehensive understanding of maser emission in such extreme environments.    

\begin{ack}
Authors thank Takumi Nagayama for his contributions to observational support for this project.  
The Nobeyama 45-m radio telescope and VERA are operated by Nobeyama Radio Observatory and Mizusawa VLBI Observatory, respectively, branches of the National Astronomical Observatory of Japan, and the National Institutes of Natural Sciences. HS is supported by JSPS KAKENHI Grant Number JP17K05388. HI, and MO is supported by JSPS KAKENHI Grant Number JP16H02167. 
\end{ack}

\end{document}